\documentclass[11pt]{article}
\usepackage{graphicx,amsmath,amsthm,mathrsfs,amssymb,float}
\usepackage{cite}
\usepackage[usenames]{color}
\usepackage{ulem}
\usepackage{authblk}
\usepackage{fontenc}

\newcommand{\ee}{\end{equation}}

\newcommand{\reff}[1]{(\ref{#1})}
\newcommand{\beq}{\begin{equation}}
\newcommand{\eeq}[1]{\label{#1}\end{equation}}
\newcommand{\beqa}{\begin{eqnarray}}
\newcommand{\eea}{\end{eqnarray}}
\newcommand{\eeqa}[1]{\label{#1}\end{eqnarray}}
\newcommand{\beg}{\begin{equation*}}
\newcommand{\eeg}{\end{equation*}}

\newcommand{\eq}{\!=\!}

\newcommand{\bsplit}{\begin{split}}
\newcommand{\esplit}{\end{split}}

\title{Formation of a condensate during charged collapse}

\author[1]{Ariel Edery\thanks{aedery@ubishops.ca}}
\author[2]{Benjamin Constantineau\thanks{benjamin.constantineau@umontreal.ca}}

\affil[1]{Department of Physics, Bishop's University, 2600 College Street, Sherbrooke, Qu\'{e}bec, Canada, J1M 1Z7.\vspace{8mm}}
\affil[2]{Universit\'{e} de Montr\'{e}al, 2900 boul. \'{E}douard-Montpetit, Montr\'{e}al, Qu\'{e}bec, Canada, H3T 1J4.}
\begin{document}
\date{}
\maketitle
\begin{abstract}
We observe a condensate forming in the interior of a black hole (BH) during numerical simulations of gravitational collapse of a massless charged (complex) scalar field. The magnitude of the scalar field in the interior tends to a non-zero constant; spontaneous breaking of gauge symmetry occurs and a condensate forms. This phenomena occurs in the presence of a BH without the standard symmetry breaking quartic potential; the breaking occurs via the dynamics of the system itself. We also observe that the scalar field in the interior rotates in the complex plane and show that it matches numerically the electric potential to within $1\%$. That a charged scalar condensate can form near the horizon of a black hole in the Abelian Higgs model without the standard symmetry breaking potential had previously been shown analytically in an explicit model involving a massive scalar field in an $AdS_4$ background. Our numerical simulation lends strong support to this finding, although in our case the scalar field is massless and the spacetime is asymptotically flat. 
  
\end{abstract}
\setcounter{page}{1}

\section{Introduction}

In this work, we consider the gravitational collapse of a complex (charged) massless scalar field to a charged black hole with asymptotically flat spacetime. Our model corresponds to scalar electrodynamics coupled to gravity (with the gauge field coupled to the scalar field in the usual fashion via the covariant derivative.) However, the standard symmetry breaking potential $V(\phi)= \mu^2 \phi^2 +\lambda \phi^4$  (with $\mu^2<0$ and $\lambda>0$) is absent.  Despite this absence, we observe numerically that the scalar field in the interior tends to a non-zero constant in contrast to the collapse of uncharged matter. Spontaneous breaking of the gauge symmetry occurs and a condensate forms. The breaking occurs via the dynamics of the system itself. In other words, in the presence of the charged black hole, it is energetically favorable for the magnitude of the scalar field to be non-zero in the interior. We also observe that the Ricci scalar at late times has a non-zero constant value inside confirming that a condensate of constant energy density has indeed formed in the interior. 

The scalar field is actually observed in the interior to rotate with time in the complex plane. From theoretical considerations, the angular speed $\omega$ should be equal to the ``electric potential" $a$ in the interior. We show that the numerical value for $\omega$ matches $a$ to within 1\%. Previous analytical work had already noted that despite the absence of the symmetry breaking potential $V(\phi)$, a charged scalar condensate can form near the horizon of a BH \cite{Gubser1, Gubser2}. This was shown explicitly in a model with negative cosmological contant and massive scalar field in an $AdS_4$ background \cite{Gubser1}. In our case, the spacetime is asymptotically flat and the scalar field is massless. Despite this difference, we observe numerically that a charged scalar condensate forms as in \cite{Gubser1}. In flat spacetime, it was already shown that spontaneous symmetry breaking (SSB) can occur without including explicitly the standard symmetry breaking potential \cite{Guendelman}. The authors discuss a system at finite charge density (i.e. finite chemical potential) where SSB occurs but is absent when the charge density is zero. They find that states of definite charge that are local minima of the energy correspond to the complex scalar field rotating with time with the angular speed $\omega$ identified with the ``chemical potential". As already stated above, the rotation of the scalar field is precisely what we also observe, with the angular speed $\omega$ identified in our case with the ``electric potential". One of the boundary conditions we impose on the charged collapse is that the electric potential is held fixed asymptotically. Thermodynamically, this corresponds to the grand canonical ensemble and the electric potential is then identified as the chemical potential. This clearly shows the connection between our black hole case and the flat spacetime case discussed in \cite{Guendelman}.     

The code for the numerical simulation, the initial conditions and the equations of motion governing the charged collapse are the same as those found in our previous work \cite{AH}, where the focus was on BH thermodynamics. Therefore, we will only provide enough information and equations in the next section so that the reader can follow in later sections the presentation of results pertaining to the formation of the condensate. For the upcoming section, the reader can always refer to \cite{AH} if they would like to see derivations or more details. We work in geometric units where $G\eq c \eq 1$. We also set Coulomb's constant to unity.  In these units, radius, mass, energy, time and charge have units of length. The coupling constant $e$ has units of inverse length. It's value is arbitrary as there is no experimental value for it in our model. We set it equal to unity and leave the scale unspecified (one can work in centimetres, metres, etc.). One unit of time is the amount of time it takes light to travel a radial distance of one unit in flat spacetime\footnote{For example, if the scale were specified to be  $3\times 10^{5}$km, then one unit of time would correspond to one second. It can be readily checked that the equations of motion do not depend on the choice of scale for $e$ i.e. they are invariant under $r\to b\,r$, $t\to b\,t$ and $e\to e/b$ where $b$ is a positive constant.}. The metric has Lorentzian signature $(-,+,+,+)$. Spacetime indices are in Greek and run from $0$ to $3$.

\section{Evolution and constraint equations}

\subsection{Metric in isotropic coordinates} 

During the numerical simulation, the metric is time-dependent and spherically symmetric. In isotropic coordinates it takes the form  
\beq
ds^2= -N(r,t)^2 dt^2 +\psi(r,t)^4 (dr^2 + r^2 d\Omega^2) \,.
\eeq{isometric}
The function $\psi(r,t)$ is often referred to as the conformal factor and the function $N(r,t)$ is called the lapse function. We assume asymptotic flatness so that $N$ and $\psi$ are both unity at infinity.  Note that the areal radius is given by $\psi^2 \,r$. The metric \reff{isometric} is convenient to use in numerical simulations of spherically symmetric collapse because the metric function $\psi$ is finite at the horizon \cite{AH,Khlebnikov, Finelli1, Finelli2, B-E,C-E,E-C}. It is also useful for the study of BH thermodynamics because the coordinate time $t$ coincides with the time measured by a clock at rest at infinity; the Lagrangian can then be identified with the negative of the free energy \cite{AH, Khlebnikov, C-E}. The Reissnner-Nordstrom (RN) BH in isotropic coordinates is given by \cite{AH, Bek1, Bek2}
\beq
ds^2=- \dfrac{\Big(1-\dfrac{M^2-Q^2}{4\,r^2}\Big)^2}{\Big(1 +\dfrac{M}{r} +\dfrac{M^2-Q^2}{4r^2}\Big)^2} \,\,dt^2 + \Big[1 +\dfrac{M}{r} +\dfrac{M^2-Q^2}{4r^2}\Big]^2 \Big( dr^2 + r^2\,d\Omega^2\Big) \,.
\eeq{IsoRN}    
The outer horizon in isotropic coordinates is situated at $r_+\eq \sqrt{M^2-Q^2}/2$. At this location, the lapse function is zero ($N(r_+)\eq0$) and the conformal factor is finite and given by $\psi(r_+) =[2+ 2 M /(M^2-Q^2)^{1/2}]^{1/2}$. For the Schwarzschild case ($Q=0$), $\psi(r_+)=2$ in agreement with the uncharged case \cite{C-E}.  The metric \reff{IsoRN} is valid only in the exterior static region. The coordinate $r$ in \reff{IsoRN} is valid in the range $r\ge r_+$ (the region $r\!<\!r_+$ does not correspond to the interior but a second covering of the exterior region). The metric for the interior of the RN BH is nonstationary \cite{E-C,CV1}. During our numerical simulation, the metric functions $N(r,t)$ and $\psi(r,t)$ appearing in  \reff{isometric} are nonstationary in the region $r\!<\!r_+$, reflecting the true nature of the interior spacetime. It is only in the exterior region $r\!>\!r_+$ that the metric \reff{isometric} approaches the static form \reff{IsoRN} at late times of the collapse. As pointed out in \cite{AH}, we do not observe the second (inner) horizon or timelike singularity associated with the RN BH. We observe only one horizon at $r_+$ (where $N(r_+)=0$) and a spacelike singularity as in the Schwarzschild case. This is in accord with the findings of previous numerical work on charged collapse \cite{Brady,Hod,Piran}. The inner horizon of the RN BH is an artifact of exact staticity (and exact spherical symmetry) \cite{Poisson} and it has been known since pioneering work in the 90's \cite{Poisson-Israel}, that the inner (Cauchy) horizon is unstable to perturbations.

\subsection{Equations of motion for matter}

For matter, we consider a complex (charged) scalar field coupled to an electromagnetic field $A_{\mu}$. The matter Lagrangian density $\mathcal{L}_{m}$ has a local $U(1)$ gauge symmetry and is given by 
\begin{equation}\label{L_Matter}
	\mathcal{L}_{m}=-\frac{1}{2}\left(\chi_{;\,\mu}+ieA_\mu\chi\right)g^{\mu\nu}\left(\overline{\chi}_{;\,\nu}-
	ieA_\nu\overline{\chi}\right)-\dfrac{1}{16\pi}F_{\mu\nu}\,F^{\mu\nu}
\end{equation}
where a semi-colon denotes covariant differentiation evaluated with metric \reff{isometric}, a bar denotes complex conjugation and $F_{\mu\nu}\!\equiv \!A_{\nu;\,\mu}-A_{\mu;\,\nu}$ is the electromagnetic field tensor. Spherical symmetry and gauge freedom reduces the number of gauge components from four to one: only $A_t=A_{0}$ is non-zero. For simplicity we label it $a$ i.e. $a=A_t$. The matter fields are therefore $\chi\eq \chi(r,t)$ and $a\eq a(r,t)$. Lagrange's equations of motion for matter are 
\begin{equation}\label{Poisson}
	\nabla_\alpha\frac{\partial\mathcal{L}_m}{\partial q_{;\,\alpha}}-\frac{\partial\mathcal{L}_m}{\partial q}=0
\end{equation}
where $q$ is a generic field.

\textit{Equations of motion for scalar field $\chi$}

We define the quantity
\begin{equation}\label{p}
	p\equiv\frac{\psi^6}{N}\left(\dot{\chi}+iea\chi\right)\,.
\end{equation}
The evolution equations for $p$ and $\chi$ respectively are \cite{AH}
\begin{equation}\label{wave}
	\dot{p}=\frac{1}{r^2}\,\partial_r\left(N\psi^2r^2\chi'\right)-ieap.
\end{equation}
and 
\begin{equation}\label{e2}
	\dot{\chi}=\frac{N}{\psi^6}\left(p-\frac{iea\psi^6\chi}{N}\right).
\end{equation} 

\textit{Equations of motion for gauge field $a$}

The constraint equation for the gauge field $a$ (one can view it as Poisson's equation) is given by
\begin{equation}\label{M1}
		\frac{1}{r^2}\partial_r\left(\frac{\psi^2a'r^2}{N}\right)=2\pi ie\,\left(\chi\,\overline{p}-\overline{\chi}\,p\right)\,.
\end{equation} 

Defining the function $g$ as 
\begin{equation}\label{g}
	g\equiv\frac{a'\psi^2}{N}\,.
\end{equation}
its evolution is given by
\begin{equation}\label{M2}
	\dot{g}=2\pi ieN\psi^2\left(\chi\,\overline{\chi}'-\overline{\chi}\,\chi'\right).
\end{equation}
The equation governing the gauge field $a$ is not an evolution equation but an ordinary differential equation: 
$a'= g\,N/ \psi^2$.

\subsection{Gravitational sector} 

\textit{Stress-energy tensor}

The stress-energy tensor $T_{\mu\nu}$ appearing in Einstein's field equations can be calculated from the matter Lagrangian \reff{L_Matter}. Its non-zero components are

\begin{equation}
	\begin{aligned}
		T_{\theta\theta} &=\dfrac{1}{2}\left(\dfrac{p\overline{p}}{\psi^{8}}-\chi'\,\overline{\chi}'
										 +\dfrac{g^2}{4\pi\psi^4}\right)r^2
		&T_{\phi\phi} 	  & =T_{\theta\theta}\sin^2\theta\\
		T_{rr} &=\dfrac{1}{2}\left(\dfrac{p\overline{p}}{\psi^{8}}+\chi'\,\overline{\chi}'
										 -\dfrac{g^2}{4\pi\psi^4}\right)
		&T_{rt}				&	 =\dfrac{N}{2\psi^6}\left(\chi'\overline{p}+\overline{\chi}'p\right)\\
		T_{tt}					 &=\dfrac{N^2}{2}\left(\frac{p\overline{p}}{\psi^{12}}+\dfrac{\chi'\overline{\chi}'}{\psi^4}
										 +\dfrac{g^2}{4\pi\,\psi^8}\right).
	\end{aligned}
\end{equation}

\textit{Field equations}

Einstein's field equations are given by $G_{\mu\nu}=\kappa^2\,T_{\mu\nu}$
where $\kappa^2\equiv8\pi\,G \eq 8\pi$ and $G_{\mu\nu}$ is the Einstein tensor evaluated with metric \reff{isometric}.   
The $G_{rr}$ equation yields evolution equations for the conformal factor $\psi$ while $G_{tt}$ and $G_{rt}$ yield constraint equations.  $G_{\theta\theta}$ yields an ordinary differential equation for the lapse function $N$ ($G_{\phi\phi}$ yields the same equation).    

Define 
\begin{equation}\label{K}
	K\equiv  -\dfrac{6 \dot{\psi}}{N\psi}\,.
\end{equation}
$K$ is the negative of the trace of the extrinsic curvature for spacelike hypersurfaces at constant time $t$. Its evolution is given by    
\begin{equation}\label{kev}
	\begin{aligned}
		\frac{\dot{K}}{N}&=\dfrac{K^2}{2}-\dfrac{6\psi'}{\psi^5}\left(\dfrac{\psi'}{\psi}+\frac{1}{r}\right)
											-\dfrac{3N'}{N\psi^4}\left(\dfrac{2\psi'}{\psi}+\dfrac{1}{r}\right)\\
											&+\dfrac{3\kappa^2}{4}\left(\dfrac{\chi'\overline{\chi}'}{\psi^4}+\dfrac{p\overline{p}}{\psi^{12}}
											-\dfrac{g^2}{4\pi \psi^8}\right)\,.
	\end{aligned}
\end{equation}

The energy constraint is given by
\begin{equation}\label{c1}
	-\dfrac{4}{\psi^5r}\left(2\psi'+r\psi''\right)
											=\dfrac{\kappa^2}{2}\left(\dfrac{\chi'\overline{\chi}'}{\psi^4}+\dfrac{p\overline{p}}{\psi^{12}}
											+\dfrac{g^2}{4\pi\psi^8}\right)
											-\dfrac{K^2}{3}
\end{equation}
while the momentum constraint is given by 

\begin{equation}\label{Grt}
	\frac{K'}{3}=\frac{\kappa^2}{4\psi^6}\left(\chi'\overline{p}+\overline{\chi}'p\right)\,.
\end{equation}

The ordinary differential equation governing the lapse function $N$ is
\begin{equation}\label{Gthetatheta}
	\dfrac{2r}{\psi^2}\partial_r\left(\frac{\psi'}{r\psi^3}\right)+\dfrac{r}{N}\partial_r\left(\frac{N'}{r\psi^4}\right)
	=-\dfrac{\kappa^2\chi'\overline{\chi}'}{\psi^4}+\dfrac{\kappa^2g^2}{4\pi\psi^8}\,.
\end{equation}

The evolution equation \reff{kev} for $K$ can be made more numerically stable by combining it with \reff{c1} to obtain
\begin{equation}
	\begin{aligned}
		\frac{\dot{K}}{N}&=K^2-\frac{6\psi'}{\psi^5}\left(\dfrac{\psi'}{\psi}+\frac{1}{r}\right)
		-\dfrac{3N'}{N\psi^4}\left(\dfrac{2\psi'}{\psi}+\dfrac{1}{r}\right)\\
		&-\dfrac{6}{r\psi^5}\left(2\psi'+r\psi''\right)-\dfrac{3\kappa^2g^2}{8\pi\psi^8}.
	\end{aligned}
\label{kev2} \end{equation}

To summarize, the evolution equations for the matter sector are \reff{e2} and \reff{p}  for the pair ($\chi$,$p$) and \reff{M2} for $g$ respectively. The evolution equations for the gravitational sector are \reff{K} and \reff{kev2} for the pair ($\psi,K$). The gauge field $a$ and lapse function $N$ obey the ordinary differential equations \reff{g} and \reff{Gthetatheta} respectively. There is one constraint equation in the matter sector, namely ``Poisson's" equation \reff{M1}. In the gravitational sector, the energy and momentum constraint are given by equations \reff{c1} and \reff{Grt} respectively. Once the initial states and boundary conditions are specified, the evolution of the fields is unique.

\subsection{Initial state and black hole formation}

\begin{figure}
	\includegraphics[scale=0.51]{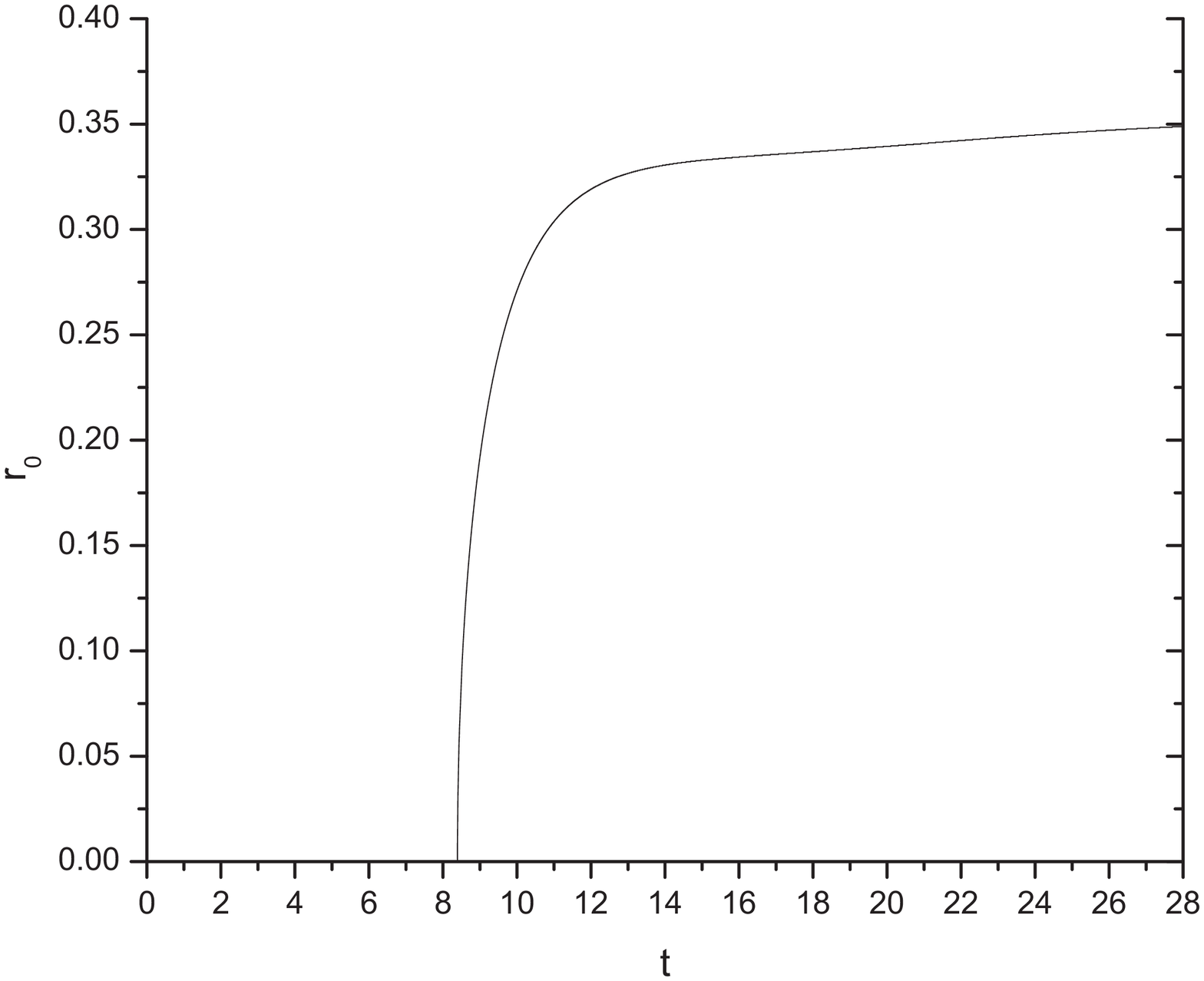}
		\caption{\label{r0} The radius $r_0$ where $N=0$ as a function of time. The lapse crosses zero for the first time at approximately t=8.4. The value of $r_0$ grows with time and approaches a value of $0.344$ at late times. }
\end{figure}

\begin{figure}
			\includegraphics[scale=0.51]{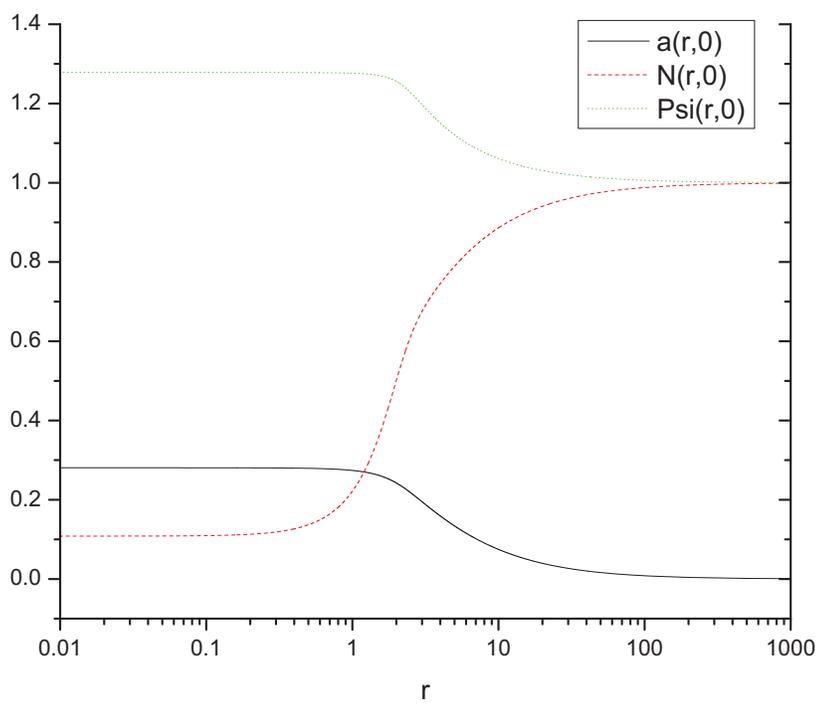}
		\caption{\label{Init} The initial states for the lapse $N$, the conformal factor $\psi$ and the ``electric potential" a. Note that the lapse is initially positive everywhere and no black hole has formed yet.}
\end{figure}

The black hole forms when the lapse function $N(r,t)$ crosses zero for the first time (this occurs at approximately t=8.4 in our simulation). The horizon grows with time i.e. the coordinate radius $r_0$ where $N=0$ grows with time and approaches at late times a value of $r_0=0.344$ (see Fig.~\ref{r0}). Initially, the lapse function is everywhere positive (see Fig.~\ref{Init}) so that there is no black hole at the start of the collapse. 

Let $\chi_1$ and $\chi_2$ be the real and imaginary part of the complex scalar field $\chi$. We begin by choosing the initial configuration for $\chi_1$, $\chi_2$, $\dot{\chi_1}$ and $\dot{\chi_2}$. We choose a kink shape given by $\chi_1=\chi_2=\lambda_2\left(\text{tanh}\left(\lambda_1-r\right)+1\right)$. Their time-derivatives are chosen to be $\dot{\chi_1}=-\dot{\chi_2}=\frac{\lambda_3\chi_1}{\lambda_2}$ (this ensures that the momentum constraint is trivially satisfied ).  One is free to choose the values of the different $\lambda_i$ and these in turn determine the values of the conserved mass $M$ and charge $Q$. For this run, we chose values of  $\lambda_1 \eq 2$, $\lambda_2 \eq 0.11$ and $\lambda_3 \eq 0.032$. The initial states for the metric functions $\psi$ and $N$ and for the gauge field $a$ are then obtained by solving three coupled second order differential equations: the energy constraint \reff{c1}, the ordinary differential equation \reff{Gthetatheta} and ``Poisson's" equation \reff{M1}. The initial state for $N$, $\psi$ and $a$ are plotted in Fig.~\ref{Init}.

\section{Numerical results: condensate formation and rotation of scalar field}


\subsection{Formation of a condensate}

\begin{figure}
		\includegraphics[scale=0.51]{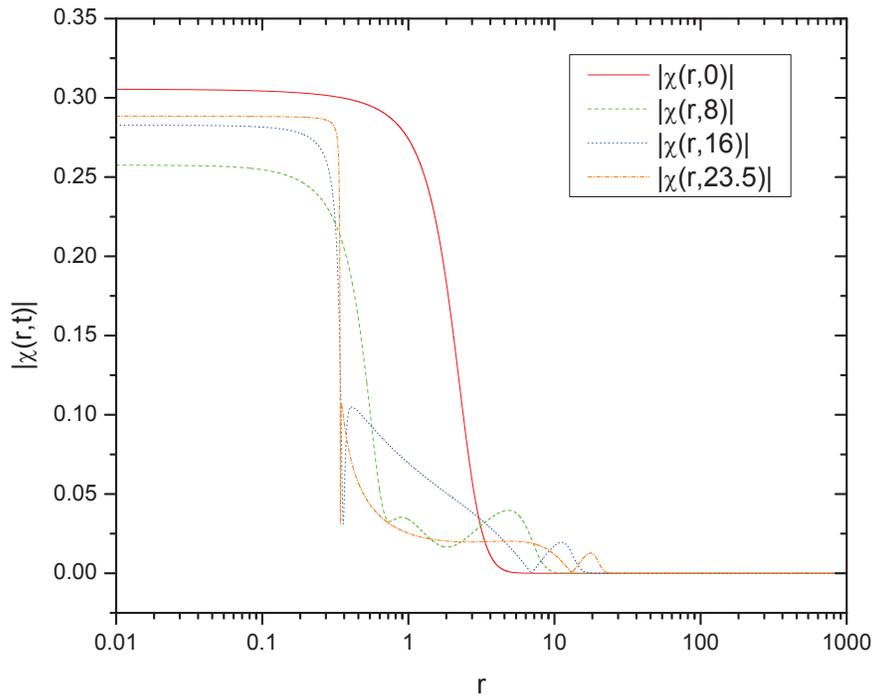}
		\caption{\label{chi2} Magnitude of charged (complex) scalar field $|\chi(r,t)|$ plotted at different times for an initial kink shape. Initially, in the interior, the magnitude decreases but then it increases and settles to a non-zero constant value at late times. This is in contrast to the uncharged case where the (real) scalar field tends to zero in the interior \cite{C-E}.}
\end{figure}

\begin{figure}[tbp]
		\begin{center}
			\includegraphics[scale=0.51, draft=false, trim=1.5cm 1.5cm 2.5cm 2cm, clip=true]{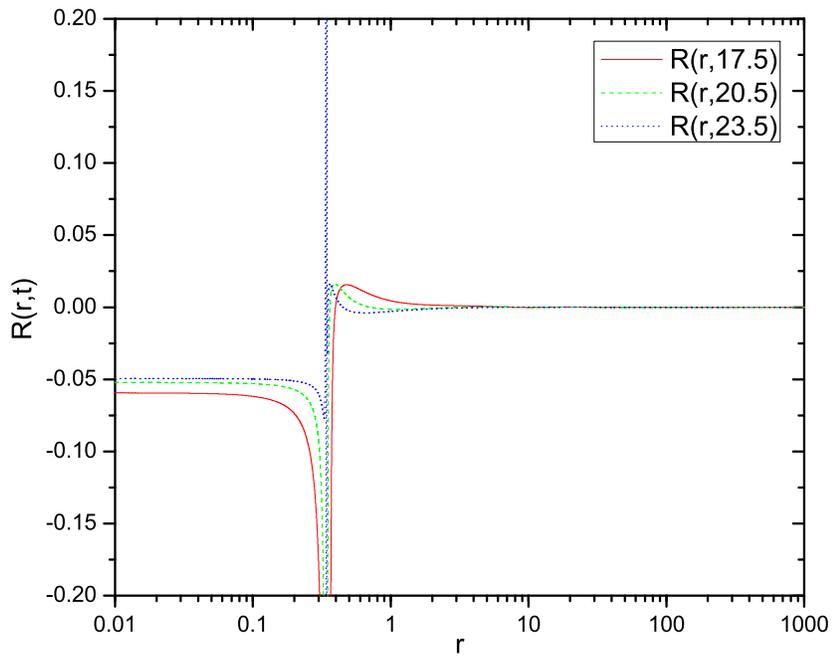}
		\end{center}
		\caption{\label{Ricci} The Ricci scalar $R(r,t)$ plotted at different times for charged collapse. It is a non-zero constant in the interior at late times signaling the presence of a constant energy density in the interior.}
\end{figure}

The magnitude of the complex (charged) scalar field at different times of the collapse is plotted in Fig.~\ref{chi2} for an initial kink shape. Initially, the field decreases in the interior but later increases to a constant value inside. At late times in the collapse process, the magnitude tends towards zero in the exterior region but has a non-zero constant value inside, signaling breaking of the Abelian gauge symmetry as was originally found analytically in \cite{Gubser1}. A condensate forms in the interior. This is in sharp contrast to the uncharged case where the value of the (real) scalar field tends to zero in the interior at late times \cite{C-E}. The Ricci scalar is plotted in Fig.~\ref{Ricci}. Note that at late times it has a non-zero constant value inside. By Einstein's field equations, the Ricci scalar is proportional to the negative of the trace of the energy momentum tensor. This implies there is a non-zero constant energy density in the interior acting much like a cosmological constant. The non-zero constant value of the Ricci scalar in the interior confirms that a condensate of constant energy density has indeed formed in the interior\footnote{There is also a Ricci singularity in the interior which is discussed in detail in \cite{B-E}.}. 

As pointed out in \cite{AH}, at late times, all aspects of the exterior RN BH are reproduced in the numerical simulation. In particular, in the exterior region, the metric functions $\psi$ and $N$ are basically static at late times and match almost exactly the RN BH metric \reff{IsoRN}. The scalar field tends towards zero in the exterior region; there is ``no scalar-hair" in agreeement with the ``no scalar-hair" conjecture for spherically symmetric, asymptotically flat static black holes \cite{Bekenstein}. The solution in \cite{Gubser1} has ``scalar-hair" but the spacetime in that case is not asymptotically flat but asymptotically $AdS_4$ and hence does not violate the ``no scalar-hair" conjecture.

\subsection{Rotation of the scalar field}

\begin{figure}
		\includegraphics[scale=0.51]{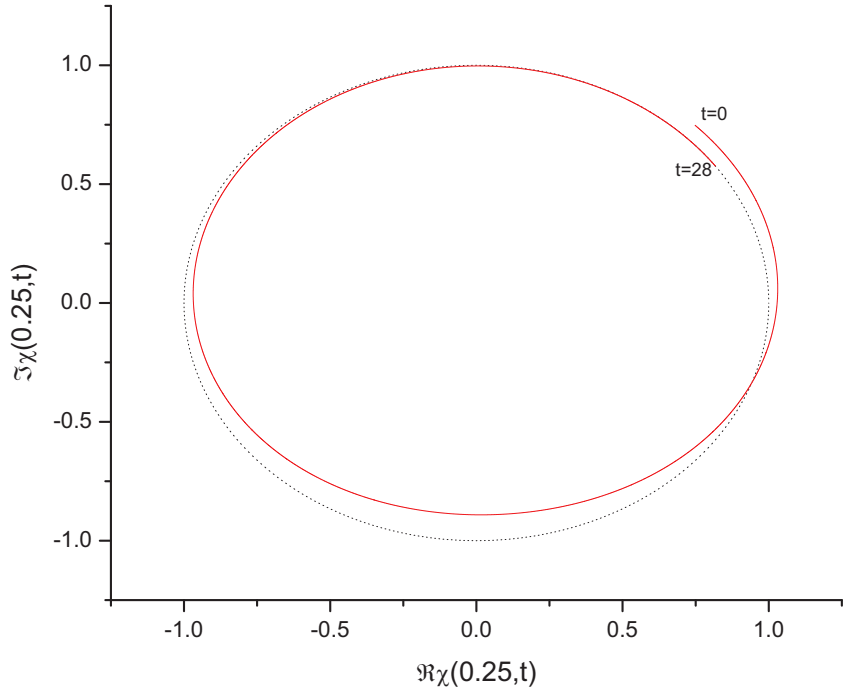}
		\caption{\label{Rotate} Plot of the real versus imaginary part of the complex scalar field taken in the interior at r=0.25 (this is the red curve). Time runs clockwise (from t=0 to t=28 in time intervals of $\delta t=0.0005$). The black curve represents the extrapolated circle constructed from points at late times. The red curve deviates from the black curve initially but matches up with the black curve at late times (i.e. approaches a circle at late times in the collapse process.) The rotation speed $\omega$ matches the average value of the electric potential $a$ at late times (see Fig.~\ref {AW})}
\end{figure}

\begin{figure}[tbp]
		\begin{center}
			\includegraphics[scale=0.51, draft=false, trim=1.5cm 1.5cm 2.5cm 2cm, clip=true]{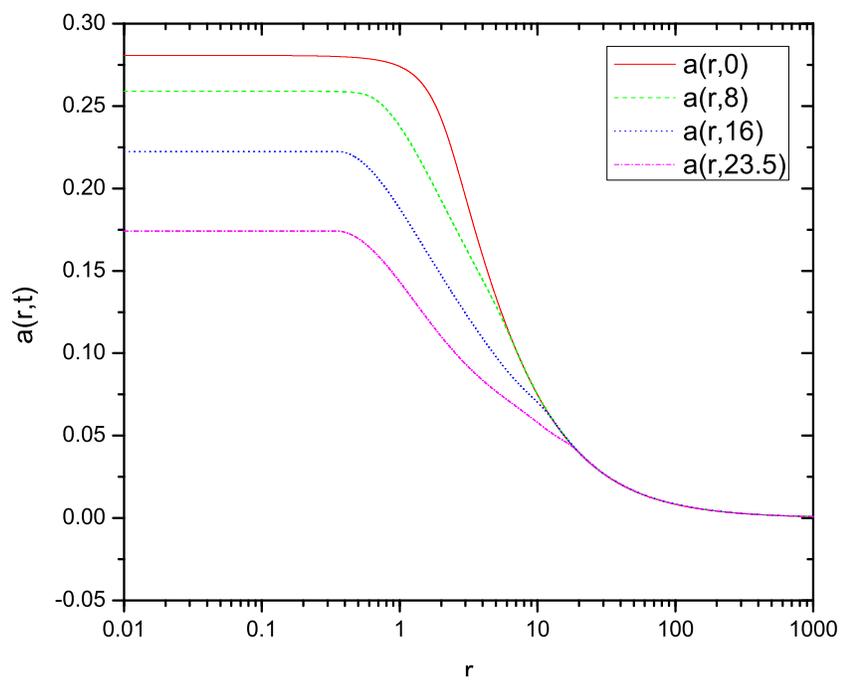}
		\end{center}
		\caption{\label{a} Potential $a(r,t)$ plotted at different times. Inside the horizon, $a$ is a non-zero constant.}
\end{figure}

\begin{figure}
		\includegraphics[scale=0.51]{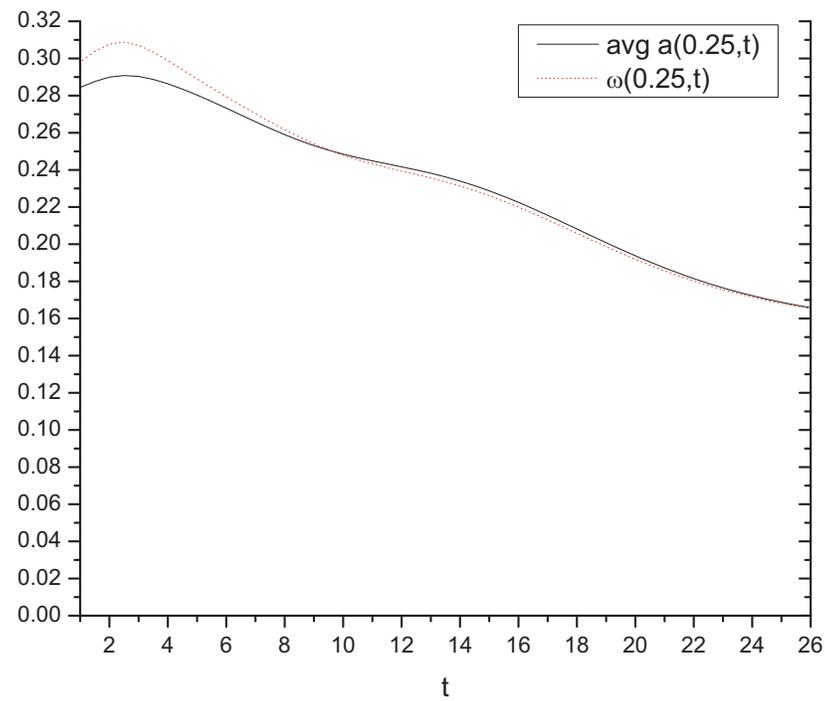}
		\caption{\label{AW} The rotation speed $\omega$ of the scalar field and the average value of the electric potential $a$ in the interior ($r=0.25$) are plotted as a function of time. Note that they approach each other with time and match at late times.}
\end{figure}

In the interior, we observe that the scalar field rotates with time in the complex plane (see Fig.~\ref{Rotate}). This is in accord with the fact that $D_{\mu}\chi=\partial_{\mu}\chi +i e A_{\mu} \chi $ approaches zero at late times in the interior (this condition minimizes the ``kinetic term" for the scalar field). This yields two equations: $\partial_{t}\chi +i e a \chi=0 $ and $\partial_{r}\chi=0$ (recall that the only non-zero component of the gauge field $A_{\mu}$ is $A_t=a$). The gauge field $a$ (which can be viewed as the ``electric potential") is set to zero at asymptotic infinity (same gauge choice as in \cite{AH})\footnote{This boundary condition corresponds thermodynamically to the grand canonical ensemble \cite{AH}.}. At late times, it approaches a non-zero constant at the horizon and the interior (see Fig.~\ref{a}). This implies that the scalar field in the interior is given by $\chi= R e^{-i\, e\, a \,t}= R e^{-i\,a\,t}$ where $R$ is the magnitude of the scalar field ($e=1$ is the value used in our numerical simulation). The scalar field therefore rotates with angular speed equal to the potential $a$ at the horizon. The coordinate $t$ in metric \reff{isometric} remains timelike in the interior; it does not switch to being spacelike as in standard coordinates and the metric functions $\psi(r,t)$ and $N(r,t)$ are continuous and finite across the horizon. Therefore, the scalar field is truly rotating with time with angular speed $a$ in the interior. The numerical data corroborates this. In Fig.~\ref{AW} we plot the rotation speed $\omega$ of the scalar field and the average value of $a$ in the interior starting at early times. Note that they converge at late times. We also made a table of the data between $t=21$ and $t=23.5$ (see table \ref{tab:table1}). The value of $\omega$ and the average value $\bar{a}$ of the electric potential agree to within less than $1\%$.  As noted in the introduction, SSB with complex scalar field rotating with time had also been found in flat spacetime in the absence of an explicit standard symmetry breaking potential \cite{Guendelman}.  
       
\subsection{Energy conditions}	

\begin{figure}
		\includegraphics[scale=0.51]{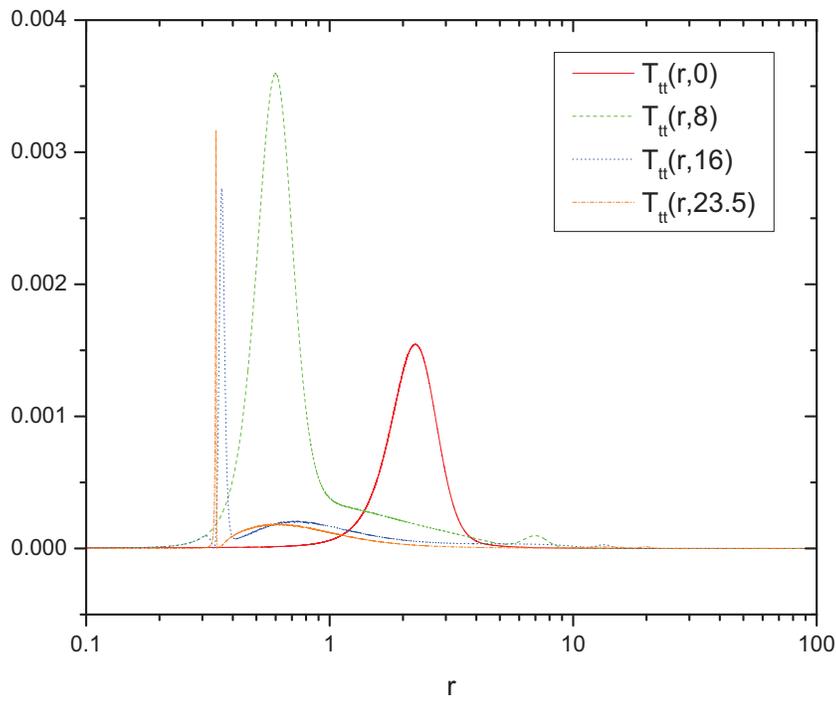}
		\caption{\label{WEC} Testing the WEC. A plot of $T_{tt}(r,t)$ at different times. Note that $T_{tt}$ is nonnegative at all times.}
\end{figure}

\begin{figure}
\centering
  \centering
  \includegraphics[scale=0.38]{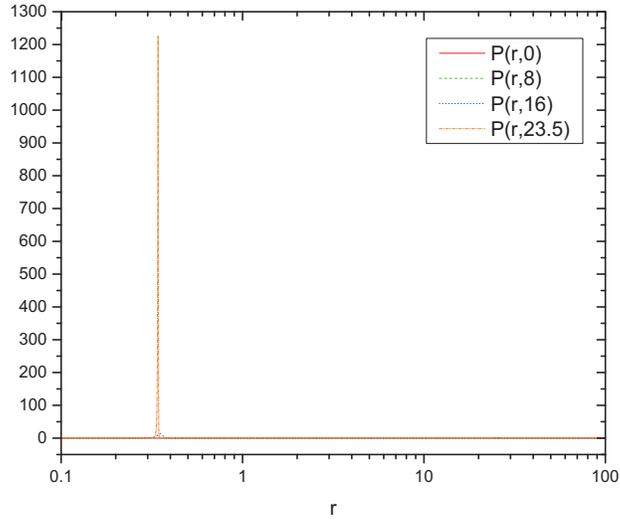}
  \caption{A plot of $P(r,t)\equiv T_{tt}/N^2 +T_{rr}/\psi^4 + 2\,T_{rt}/(N\,\psi^2)$ at different times. See figure below for a magnified version.}
  \label{NEC1}
\end{figure}

\begin{figure}
  \centering
  \includegraphics[scale=0.38]{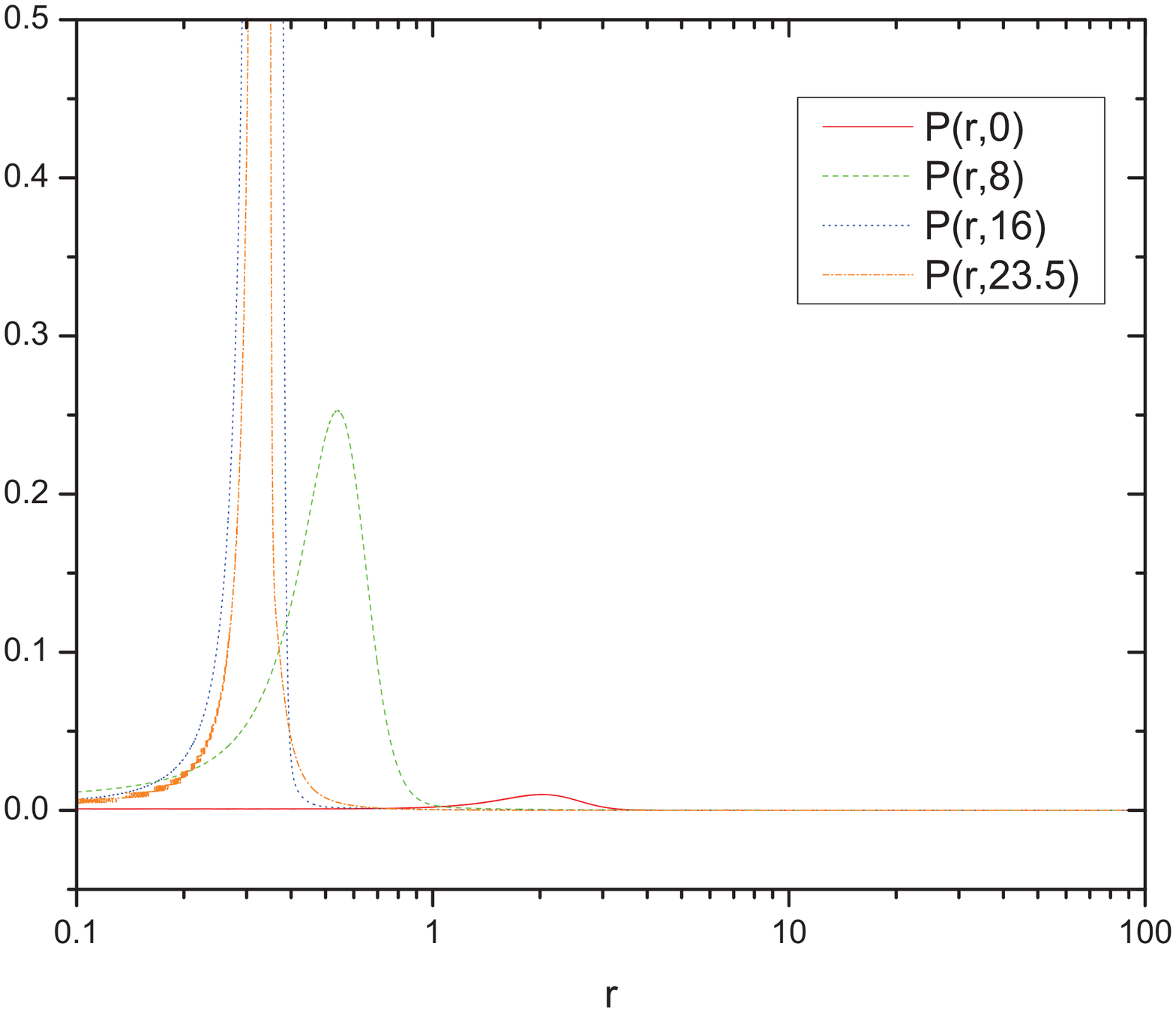}
  \caption{Testing the NEC. Note that $P$ is nonengative at all times.}
  \label{NEC2}
\end{figure}

It is common for scalar fields to violate some energy conditions. Here we will check the weak energy condition (WEC) and null energy condition (NEC) and we will see that they are both obeyed during the collapse process. The WEC requires that $T_{\mu\nu} \,t^{\mu}t^{\nu} \ge 0$ for any timelike vector $t^{\mu}$. We choose the timelike vector $t^{\mu}=(1,0,0,0)$ so that the WEC for that vector becomes $T_{tt}\ge 0$. We plot $T_{tt}(r,t)$ at different times in Fig.~\ref{WEC}. It is nonnegative at all times and the WEC is obeyed. The NEC requires that $T_{\mu\nu} \,\ell^{\mu}\ell^{\nu} \ge 0$ for any null vector $\ell^{\mu}$. We choose the null vector $\ell^{\mu} =(1/N, 1/\psi^2, 0,0)$  so that the NEC for that vector becomes $P\equiv T_{tt}/N^2 + T_{rr}/\psi^4 + 2 T_{rt}/(N\,\psi^2) \ge 0$. We plot $P(r,t)$ at different times in figures \ref{NEC1} and \ref{NEC2} (Fig.~\ref{NEC2} is a magnified version of Fig.~\ref{NEC1}).  One can see in Fig.~\ref{NEC2} that $P$ is nonnegative at all times and the NEC is obeyed. We therefore have that the NEC and WEC are both obeyed. 		
			
\begin{table}
	\centering
	\begin{tabular}{c|cccccc}\hline
		Time 				& 20.999-21 & 21.499-21.5 & 21.999-22 & 22.499-22.5 & 22.999-23 & 23.499-23.5 \\ \hline
		$\omega$ 	& 0.18570 & 0.18287 & 0.18022 & 0.17776 & 0.17549 & 0.17340 \\
		$\bar{a}$ 	& 0.18726 & 0.18428 & 0.18149 & 0.17889 & 0.17649 & 0.17429 \\
		\% diff. 			& 0.83307 & 0.76514 & 0.69976 & 0.63167 & 0.56660 & 0.510643 \\ \hline
	\end{tabular}
	\caption{The angular speed $\omega$ at which the complex scalar field rotates is compared to the average value $\bar{a}$ of the electric potential at a point in the interior (at $r=0.25$). The difference is less than 1\%.}
	\label{tab:table1}
	\end{table}	


\section{Conclusion}

We have shown that numerical simulations of charged collapse reveal that a condensate can form in the interior of a charged BH in scalar electrodynamics. Breaking of the Abelian gauge symmetry occurs in the presence of the charged BH in the absence of the standard symmetry breaking potential $V(\phi)= \mu^2 \phi^2 +\lambda \phi^4$. That a condensate can form near the horizon without the potential $V(\phi)$ had been previously found analytically for the case of a massive scalar field in an asymptotically $AdS_4$ spacetime \cite{Gubser1}. Our numerical solution lends strong support to this finding even though in our case the scalar field is massless and the spacetime is asymptotically flat. That SSB can occur without the standard potential had also been noted in flat spacetime \cite{Guendelman}. Moreover, as in our case, they obtain that the complex scalar field rotates with time. This clearly demonstrates that the phenomena of SSB can occur by the properties of the system itself, without an explicit potential, in either flat or curved spacetime.

The standard stationary BH solutions like Schwarzschild, RN and Kerr do not include a scalar field to model uncharged or charged matter. For Schwarzschild, the standard solution yields a Ricci tensor and Ricci scalar that is zero everywhere including the interior since this is a vacuum solution.  However, in spherically symmetric gravitational collapse of uncharged matter, the Ricci tensor and Ricci scalar will not be zero inside. In fact, we expect a Ricci singularity in the interior where the energy density diverges. The interiors of the standard and gravitational collapse solutions clearly do not match; there is an exact match with the standard Schwarzschild solution only in the exterior region. In the stationary RN BH case, the standard solution does not include a complex scalar field or Dirac field to model the charged matter. It is therefore not surprising that the gravitational collapse  of a charged scalar field can provide something novel {\it in the interior} that was not revealed by the standard solution.  Again, the numerical solution for gravitational collapse matches the exterior region of the standard RN solution exactly. In particular, the Ricci scalar is zero in the exterior region as in the standard solution.   

What is of course remarkable is that despite their different interiors, the gravitational collapse \cite{Khlebnikov, AH, C-E, Brady, Hod, Piran} and standard solutions have the same BH entropy and thermodynamics for the same conserved quantities (such as the mass $M$ and charge $Q$). As stated in \cite{Bekenstein}, this is intimately connected to the ``no scalar hair" theorem. This theorem however depends on having asymptotic flatness. It would therefore be of great interest to investigate numerically what happens in the collapse of a charged scalar field to a BH in an asymptotically curved background (such as de Sitter or anti-de Sitter space). In particular, one could investigate the thermodynamics of such BH's using the Lagrangian as a probe into the free energy as was recently done \cite{Khlebnikov, AH, C-E} and determine if there is a deviation from standard BH thermodynamics due to the presence of the condensate.

\section*{Acknowledgments}
A.E. acknowledges support from a discovery grant of the National Science and Engineering Research Council of Canada (NSERC). AE thanks Yu Nakayama for discussions and for pointing out reference 1. We thank Hugues Beauchesne for making the charged code freely available to us.



\begin{thebibliography}{99}
\bibitem{Gubser1} S. Gubser, Phys. Rev. D {\bf 78}, 65034 (2008) [arXiv:0801.2977].
\bibitem{Gubser2} S. Gubser, Class. Quantum Grav. {\bf 22}, 5121 (2005) [hep-th/0505189].
\bibitem{Guendelman} J. Bekenstein and E. Guendelman, Phys. Rev. D {\bf 35}, 716 (1987). 
\bibitem{AH} H. Beauchesne and A. Edery, JHEP {\bf 05}, 146 (2012) [arXiv:1203.2279]
\bibitem{Khlebnikov} Z. Gecse and S. Khlebnikov, Phys. Rev. D {\bf 77}, 104003 (2008) [arXiv:0801.3662].
\bibitem{Finelli1} F. Finelli, and S. Khlebnikov, Phys. Lett. B {\bf 504}, 309 (2001).
\bibitem{Finelli2} F. Finelli and S. Khlebnikov, Phys. Rev. D {\bf 65}, 043505 (2002).
\bibitem{B-E} H. Beauchesne and A. Edery, Phys. Rev. D {\bf 85}, 044056 (2012) [arXiv:1108.0449].
\bibitem{C-E} B. Constantineau and A. Edery, 	Phys. Rev. D {\bf 84}, 084032 (2011) [arXiv:1103.5272].
\bibitem{E-C} A. Edery and B. Constantineau, Class. Quantum Grav. {\bf 28}, 045003 (2011) 
\bibitem{Bek1} J. Bekenstein and M. Schiffer, Phys. Rev. D {\bf 80}, 123508 (2009) [arXiv:0906.4557].
\bibitem{Bek2} E. Sagi and J. Bekenstein, Phys. Rev. D {\bf 77}, 024010 (2008) [arXiv:0708.2639].
\bibitem{CV1} C. Vaz and L. Witten, Phys. Rev. D {\bf 64}, 084005 (2001).
\bibitem{Poisson} E. Poisson, {\it A Relativist's Toolkit} (Cambridge University Press, Cambridge, 2004).
\bibitem{Brady} P. Brady and J. Smith, Phys. Rev. Lett. {\bf 75}, 1256 (1995) [arXiv:gr-qc/9506067]. 
\bibitem{Hod} S. Hod and T. Piran, Phys. Rev. Lett. {\bf 81}, 1554 (1998).
\bibitem{Piran} Y. Oren and T. Piran, Phys. Rev. D {\bf 68}, 044013 (2003).
\bibitem{Poisson-Israel} E. Poisson and W. Israel, Phys. Rev. D {\bf 41}, 1796 (1990).
\bibitem{Bekenstein} J. Bekenstein, {\it Proceedings of the second Sakharov conference in Physics, Moscow, 1996.} arXiv:gr-qc/9605059. 
[arXiv:1010.5844].
\end{thebibliography}
\end{document}